\documentclass[12pt]{article}
\usepackage{axodraw,graphicx,psfrag,amssymb,cite}

\renewcommand{\thefootnote}{\fnsymbol{footnote}}

\newcommand{\prepr}[1] {\begin{flushright}  {\bf #1} \end{flushright} \vskip 1.cm}
\newcommand{\titul}[1] {\boldmath \begin{center}{\Large {\bf #1 } } \end{center}
\vskip 0.8cm}

\newcommand{\autor}[1] {\begin{center}  {\bf \lineskip .3cm #1  }
                        \end{center} }

\newcommand{\lugar}[1] {\begin{center}  {\normalsize \bf \it #1   } \end{center}}
\newcounter{muni}

\makeatletter
\def\fmslash{\@ifnextchar[{\fmsl@sh}{\fmsl@sh[0mu]}}
\def\fmsl@sh[#1]#2{%
  \mathchoice
    {\@fmsl@sh\displaystyle{#1}{#2}}%
    {\@fmsl@sh\textstyle{#1}{#2}}%
    {\@fmsl@sh\scriptstyle{#1}{#2}}%
    {\@fmsl@sh\scriptscriptstyle{#1}{#2}}}
\def\@fmsl@sh#1#2#3{\m@th\ooalign{$\hfil#1\mkern#2/\hfil$\crcr$#1#3$}}
\makeatother
\begin{document}
\hbadness=10000
\pagenumbering{arabic}
\begin{titlepage}

\prepr{hep-ph/yymmnnn\\
\hspace{30mm} January 2007}

\begin{center}
\titul{\bf Effect of $H^\pm$ on $B^{\pm}\to \tau^\pm\nu_\tau$ 
and $D^\pm_s\to \mu^\pm\nu_\mu,\tau^\pm\nu_\tau$}

\autor{A.G. Akeroyd\footnote{akeroyd@mail.ncku.edu.tw} and 
Chuan Hung Chen\footnote{physchen@mail.ncku.edu.tw}}
\lugar{1: Department of Physics,\\
National Cheng Kung University,\\
Tainan 701, Taiwan}
\lugar{2: National Center for Theoretical Sciences,\\
Taiwan}
\end{center}

\vskip2.0cm

\begin{abstract}
\noindent{The recent observation of the purely leptonic
decay $B^\pm\to \tau^\pm\nu_\tau$ at the
B factories permits a sizeable contribution from a charged
Higgs boson ($H^{\pm}$). Such a $H^{\pm}$ would also contribute to 
the decays $D^\pm_s\to \mu^\pm\nu_{\mu}$ and 
$D^\pm_s\to \tau^\pm\nu_{\tau}$, which 
are being measured with increasing precision at CLEO-c. 
We show that the branching ratios of $D^\pm_s\to \mu^\pm\nu_{\mu}$ and 
$D^\pm_s\to \tau^\pm\nu_{\tau}$ could be suppressed
by up to $10\%$ from the Standard Model prediction,
which is larger than the anticipated precision in 
the measurements of these decays at forthcoming BES-III. 
}

\end{abstract}

\vskip1.0cm
\vskip1.0cm
{\bf Keywords : \small Charged Higgs boson, Rare B decay, Rare D decay} 
\end{titlepage}
\thispagestyle{empty}
\newpage

\pagestyle{plain}
\renewcommand{\thefootnote}{\arabic{footnote} }
\setcounter{footnote}{0}

\section{Introduction}

In April 2006 the BELLE collaboration announced the first observation of the
purely leptonic decay $B^\pm\to \tau^\pm \nu_\tau$ \cite{Ikado:2006un}
utilizing an integrated luminosity of $414$ fb$^{-1}$. The measured
branching ratio (BR) is in agreement with the Standard Model (SM) rate
within theoretical and experimental errors:
\begin{equation}
{\rm BR}(B^\pm\to \tau^\pm\nu_\tau)=
(1.79^{+0.56}_{-0.49}(stat)^{+0.46}_{-0.51}(syst))
\times 10^{-4}
\label{exp_btaunu}
\end{equation}
Subsequently, BABAR reported an improved upper limit with 
288 fb$^{-1}$ \cite{Aubert:2006fk}:
\begin{equation}
{\rm BR}(B^\pm\to \tau^\pm\nu_\tau) < 1.8\times 10^{-4} \,\,90\%\, c.l
\end{equation}
Significantly improved precision for BR$(B^\pm\to \tau^\pm\nu_\tau)$
would require a high luminosity upgrade of the
existing B factories \cite{Yamauchi:2002ru}.
Such a facility would also provide sensitivity to the
SM rate for the analogous leptonic decay $B^\pm\to \mu^\pm\nu_\mu$.

In the SM the decay $B^\pm\to \tau^\pm\nu_\tau$ proceeds via
annihilation of $B^\pm$ to a $W^*$.
It has been known for some time that New Physics could significantly
affect BR$(B^\pm\to \tau^\pm\nu_\tau)$. 
A charged Higgs boson ($H^\pm$) present in the Two Higgs Doublet Model (2HDM)
would also mediate the annihilation of $B^\pm$ \cite{Hou:1992sy}, 
and for Model II type Yukawa couplings
its effect on BR$(B^\pm\to \tau^\pm\nu_\tau)$ is determined by
the ratio $R=\tan\beta/m_H$ (where $m_H$ is the mass
of $H^\pm$, $\tan\beta=v_2/v_1$ and $v_1,v_2$
are vacuum expectation values of the Higgs doublets).
The contribution from $H^{\pm}$ interferes 
{\sl destructively} with the $W^\pm$ mediated SM diagram, 
and thus the experimental measurement of 
BR($B^\pm\to \tau^\pm\nu_\tau$) leads to two allowable regions
for $R$ \cite{Ikadotalk}:
\begin{itemize}
\item[{(i)}] A region where the $H^\pm$ contribution 
is small (a perturbation to the SM rate), roughly 
corresponding to $0< R <0.15$ GeV$^{-1}$.
\item[{(ii)}] A region where $H^\pm$ contributes sizeably but
the large destructive interference maintains the SM rate, 
roughly corresponding to $0.22$ GeV$^{-1}$ $< R < 0.33$ GeV$^{-1}$. 
\end{itemize}

In a more general 2HDM in which the bottom quark receives mass
from both Higgs doublets (e.g. which includes the 
Minimal Supersymmetric SM at the one-loop level) 
the above regions for $R$ are altered \cite{Akeroyd:2003zr}.
Of particular interest is region (ii) since such 
sizeable values of $R$ might have an observable effect on other processes
in which $H^{\pm}$ contributes at tree-level.

The hitherto unobserved exclusive semi-leptonic decay 
$B\to D\tau^\pm \nu_\tau$ (where $B$ can be charged or neutral)
has a relatively large BR $(\approx 8\times 10^{-3})$
in the SM, and can also be mediated by $H^{\pm}$ 
\cite{Tanaka:1994ay},\cite{Itoh:2004ye}.
A first observation of this decay would require several ab$^{-1}$
of integrated luminosity.
The angular distribution of the $\tau$ \cite{Chen:2006nu} would 
also provide information on $R$ with reduced 
uncertainties from Cabbibo-Kobayashi-Maskawa (CKM)
matrix elements and strong interaction effects.
Direct production of $H^\pm$ at the Tevatron (via $t\to H^\pm b$)
\cite{Abulencia:2005jd} and 
at the LHC (e.g. via $gb\to t H^\pm$)
would also probe the region (ii). Notably, 
LHC simulations \cite{Assamagan:2002ne}
promise verification or falsification of region (ii) for
$\tan\beta< 100$, as well as probing a sizeable fraction of region (i).

In this paper we consider the effect of $H^\pm$ on the purely leptonic 
decays of the
charmed meson $D^\pm_s$. Due to the smaller Yukawa couplings
of $H^\pm$ to the lighter quarks the effect of $H^\pm$ on
such decays will clearly be smaller than 
the effect on $B^\pm\to \tau^\pm\nu_\tau$.
However,  BR$(D^\pm_s\to \mu^\pm\nu_\mu)\sim 5\times 10^{-3}$ and 
BR$(D^\pm_s\to \tau^\pm\nu_\tau)\sim 5\times 10^{-2}$
are much larger than BR($B^{\pm}\to \tau^\pm\nu_\tau$) and thus
small perturbations to the SM rate from $H^\pm$ 
might be accessible to ongoing and forthcoming experiments.
Measurements of BR$(D^\pm_s\to \mu^\pm\nu_\mu)$ and 
BR$(D^\pm_s\to \tau^\pm\nu_\tau)$
have been performed by a variety of experiments
(for a summary see \cite{Soldner-Rembold:2001zk}) and are being improved by 
ongoing CLEO-c \cite{Artuso:2006kz},\cite{Stone:2006cc}.
 Moreover, BABAR 
has recently observed $D^\pm_s\to \mu^\pm\nu_\mu$ \cite{Aubert:2006sd},
with a precision comparable to that at CLEO-c. At BES-III,
which is due to commence in late 2007, a
precision of $2\%$ for BR$(D^\pm_s\to \mu^\pm\nu_\mu)$ and 
BR$(D^\pm_s\to \tau^\pm\nu_\tau)$ is expected
after 4 years of operation at design luminosity \cite{Li:2006nv}.

To date, the main motivation for searching for
$D^\pm_s\to \mu^\pm\nu_\mu$ and $D^\pm_s\to \tau^\pm\nu_\tau$
has been to measure the decay constant
$f_{D_s}$, assuming that no New Physics contributes. 
Such measurements are vital tests of lattice predictions for the
heavy quark systems.
However, if $H^\pm$ is contributing sizeably to 
$B^\pm\to \tau^\pm\nu_\tau$ it would have a non-negligible effect on  
$D^\pm_s\to \mu^\pm\nu_{\mu}$ and $D^\pm_s\to\tau^\pm\nu_{\tau}$.
This was noted in \cite{Hou:1992sy},\cite{Hewett:1995aw} 
and a first quantitative study was performed
in \cite{Akeroyd:2003jb}. Motivated by the recent observation
of $B^\pm\to \tau^\pm\nu_\tau$, in this paper we update and extend the
analysis of \cite{Akeroyd:2003jb}, assuming that $H^\pm$ 
is contributing sizeably to $B^\pm\to \tau^\pm\nu_\tau$.
We point out that a precise measurement of the 
ratio BR($D^\pm_s\to \mu^\pm\nu_{\mu}$)/BR($D^\pm\to \mu^\pm\nu_{\mu}$)
(which will be possible at BES-III) could provide a hint for the
presence of $H^\pm$. Our work is organized as follows.
In section 2 we review the contribution of 
$H^\pm$ to the leptonic decays and summarize the current
experimental situation. The numerical analysis
is contained in section 3 with conclusions given in
section 4.

\section{The decays $B^\pm\to l^\pm\nu_l$ and $D^\pm_{(s)}\to l^\pm\nu_l$}
\begin{figure}
\begin{center}
\includegraphics[width=6.5cm,angle=0]{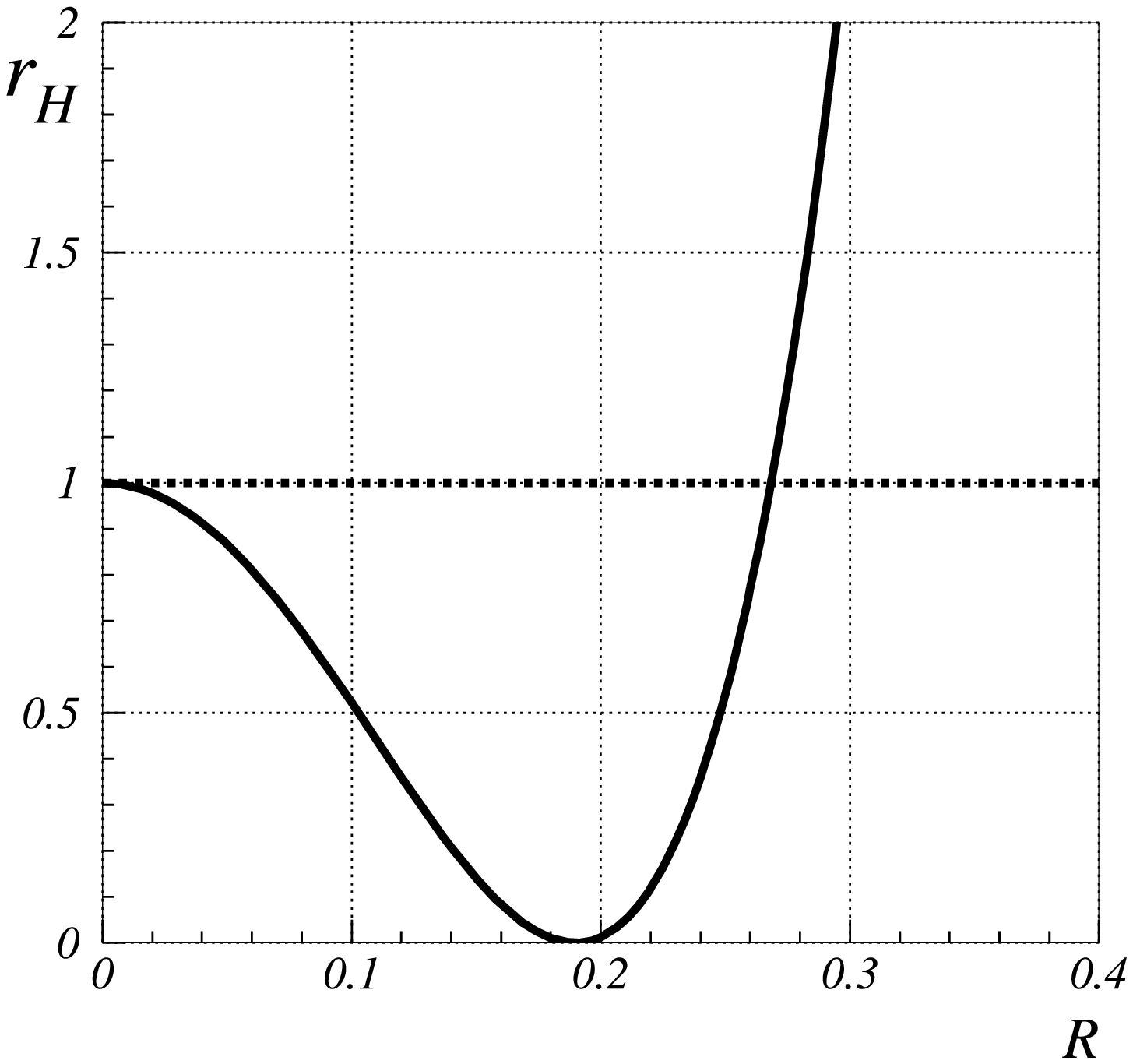} 
\includegraphics[width=6.5cm,angle=0]{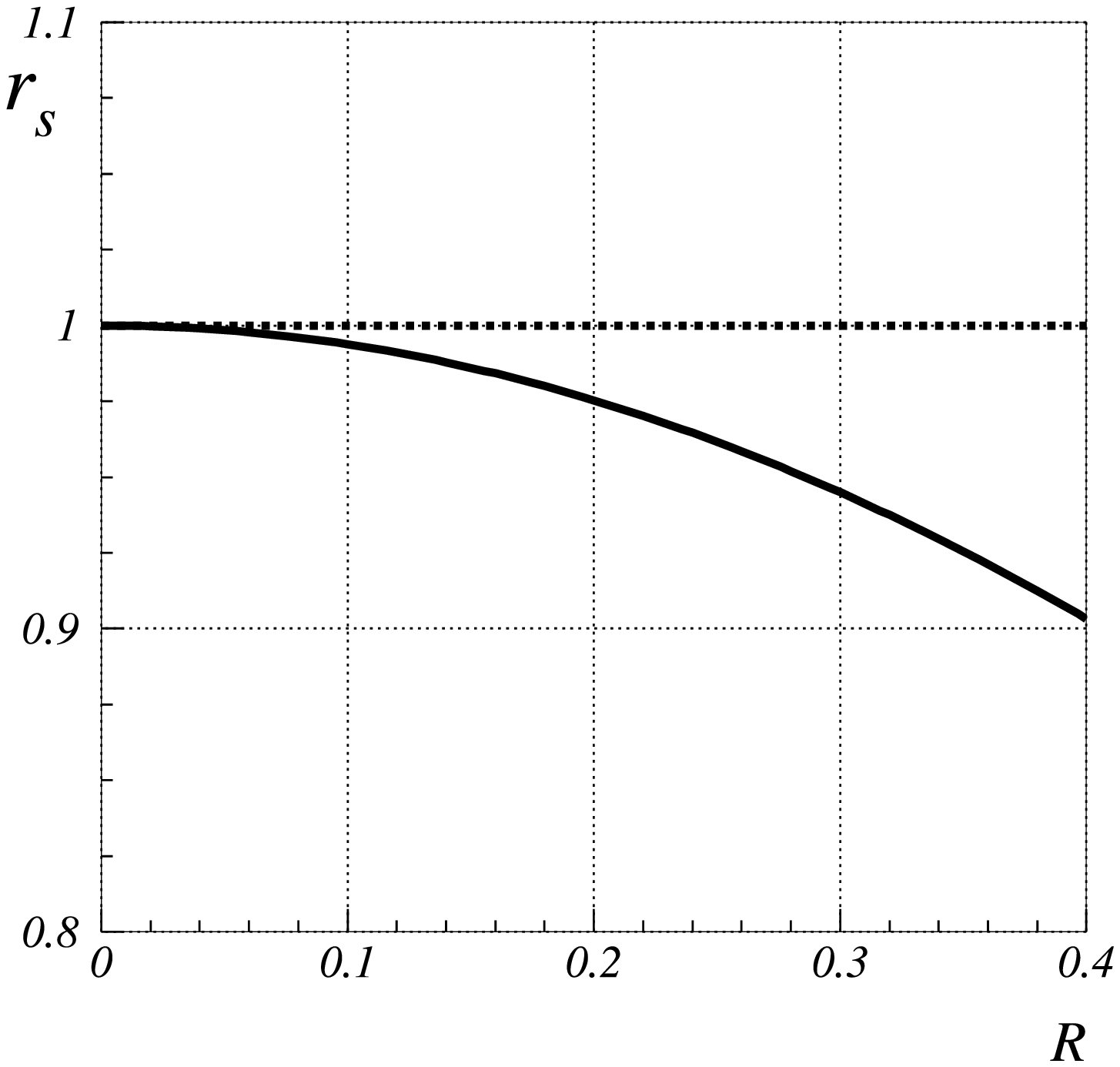}
\caption{Scaling factor of the SM rate (solid line) 
as a function of $R(=\tan\beta/m_{H^\pm})$.
Left panel (Fig.1a): $r_H$ for $B^\pm\to \tau^\pm\nu_{\tau}$; 
Right panel (Fig.1b): $r_s$ for
$D^\pm_s\to \mu^\pm\nu_\mu,\tau^\pm\nu_{\tau}$ with $m_s/(m_c+m_s)=0.08$.}
\label{fig1}
\end{center}
\end{figure}

In the SM the purely leptonic decays $B^\pm\to l^\pm\nu_l$ and 
$D^\pm_{(s)}\to l^\pm\nu_l$ proceed via annihilation of the 
heavy meson
into $W^*$. Singly charged Higgs bosons,
which arise in any extension of the SM with
at least two $SU(2)_L\times U(1)_Y$ Higgs doublets with hypercharge
$Y=1$, would also contribute to these decays.
For $B^-\to l^-\overline \nu_l$ the Feynman diagram is as follows:
\begin{center}
\vspace{-50pt} \hfill \\
\begin{picture}(120,90)(0,25) 
\Photon(10,25)(68,25){4}{8}
\Vertex(10,25){3}
\ArrowLine(-40,55)(10,25)
\ArrowLine(10,25)(-40,-5)
\ArrowLine(68,25)(118,-5)
\ArrowLine(118,55)(68,25)
\Text(-18,50)[]{$b$}
\Text(-13,0)[]{$\overline u$}
\Text(135,-2)[]{$l^-$}
\Text(129,55)[]{$\overline \nu_l$}
\Text(41,38)[]{$W^-,H^-$}
\GOval(-45,25)(33,10)(0){0.5}
\Text(-70,25)[]{$B^-$}
\end{picture}
\end{center}
\vskip 1cm
The tree--level partial width is given by \cite{Hou:1992sy}:
\begin{equation}
\Gamma(B^\pm\to l^\pm\nu_l)=\frac{G_F^2 m_{B} m_l^2 f_{B}^2}{8\pi}
|V_{ub}|^2 \left(1-\frac{m_l^2}{m^2_{B}}\right)^2 \,\times\, r_H
\end{equation}
where $G_F$ is the Fermi constant,  
$m_l$ is the mass of the lepton, $m_B$ is the mass of the
$B$ meson, $V_{ub}$ is a CKM matrix element, and $f_B$ is the
decay constant. In the 2HDM (Model II), in which the $b$ quark
only couples to one of Higgs doublets at tree-level, the scaling
factor $r_H$ of the SM rate is given by:  
\begin{equation}
r_H=[1-m^2_B\frac{\tan^2\beta}{m^2_{H^\pm}}]^2\equiv [1-m_B^2R^2]^2.
\label{r_H}
\end{equation}
The $H^\pm$ contribution interferes destructively with 
that of $W^\pm$. There are two solutions for 
$r_H=1$ which occur at $R=0$ and $R=0.27$ GeV$^{-1}$
(see Fig.\ref{fig1}a). 
If the $b$ quark couples to both Higgs doublets at tree-level,
which is referred to as the 2HDM (Model III) \cite{Lee:1973iz},
Eq.~(\ref{r_H}) is modified to \cite{Akeroyd:2003zr}:
\begin{equation}
r_H=\left(1-{\tan^2\beta\over 1+\tilde\epsilon_0\,\tan\beta}\,
\frac{m^2_B} {m^2_{H^\pm}}\right)^2
\label{mod_r_H}
\end{equation}
In the Minimal Supersymmetric SM (MSSM)
the parameter $\tilde\epsilon_0$ is generated at 
the 1-loop level  \cite{Banks:1987iu},\cite{Hall:1993gn}
(with the main contribution originating from gluino diagrams)
and may reach values of 0.01. 
The redefinition of both the $b$ quark 
Yukawa coupling and the CKM matrix element  
$V_{ub}$ are encoded in $\tilde\epsilon_0$ \cite{D'Ambrosio:2002ex}. 
The impact of $\tilde\epsilon_0\ne 0$ 
on $r_H$ has been developed in \cite{Itoh:2004ye},
\cite{Chen:2006nu},\cite{Isidori:2006pk},
\cite{Idarraga:2005ia}.
In particular, the value of $R$ where $r_H=0$ and $r_H=1$ shifts
depending on the magnitude and sign of $\tilde\epsilon_0$.

For the decay $D^-_{(s)}\to l^-\overline\nu_{l}$ 
the partial width is given by \cite{Hou:1992sy}:
\begin{equation}
\Gamma(D^\pm_{(s)}\to l^\pm\nu_l)=\frac{G_F^2 m_{D_{(s)}} 
m_l^2 f_{D_{(s)}}^2}{8\pi}|V_{cd(cs)}|^2 
\left(1-\frac{m_l^2}{m^2_{D_{(s)}}}\right)^2\times r_{(s)}
\end{equation}
where $m_{D_{(s)}}$ is the mass of the
$D^\pm_{(s)}$ meson, $V_{cd(cs)}$ are CKM matrix elements,
$f_{D_{(s)}}$ are decay constants, and
$r_{(s)}$ is the analogy of $r_H$ given by:
\begin{equation}
r_{(s)}=[1-m^2_{D_q}\frac{\tan^2\beta}{m^2_{H^\pm}}
(\frac{m_q}{m_c+m_q})]^2\equiv
[1-m^2_{D_q}R^2(\frac{m_q}{m_c+m_q})]^2
\label{r_s}
\end{equation}
As in the case for $B^\pm\to l^\pm\nu_l$, the $H^\pm$ contribution depends
on $R=\tan\beta/m_{H^\pm}$, although the factor  
$m^2_{D_q}m_q/(m_c+m_q)$ in Eq~.(\ref{r_s}) is considerably smaller than
$m_B$ present in Eq~.(\ref{r_H}).
The $H^\pm$ contribution to $D^\pm\to l^\pm\nu_l$ is essentially negligible 
due to the smallness of
$m_d/m_c$ and thus $r\approx 1$. However,
for $D^\pm_s$ the scaling factor $r_s$ may differ from 1 due to the
non--negligible $m_s/m_c$
\cite{Hou:1992sy},\cite{Hewett:1995aw},\cite{Akeroyd:2003jb}.
The expression for $r_s$ in Eq.~(\ref{r_s}) can contain an
$\tilde\epsilon_0$ correction analogous to Eq.~(\ref{mod_r_H}),
but we do not write this contribution explicitly due to
the considerable uncertainty in $m_s/(m_c+m_s)$.
In Fig.\ref{fig1}b we plot $r_s$ as a function of $R$.
Although the effect of $H^\pm$ is only a perturbation to the SM rate for
BR$(D^\pm_s\to \mu^\pm\nu_\mu,\tau^\pm\nu_\tau)$, such deviations
might be measurable since their BRs in the SM are much larger
than BR$(B^\pm\to \tau^\pm\nu_\tau)$.
Prospects for precise measurements of 
BR($D^\pm_s\to \mu^\pm\nu_\mu,\tau^\pm\nu_\tau$) are bright. The ongoing
CLEO-c programme \cite{Shipsey:2002ye}, \cite{Poling:2006da} 
(which will terminate around April 2008)
utilizes the production mechanism $e^+e^-\to D_s^{\pm *}D_s^\pm$,
and expects a final precision $\sim 10\%$ for
BR($D^\pm_s\to \mu^\pm\nu_\mu,\tau^\pm\nu_\tau$).
At forthcoming BES-III
an integrated luminosity of 20 fb$^{-1}$ (which corresponds 
to 4 years at design luminosity) would enable
a precision $\sim 2\%$ for BR($D^\pm_s\to \mu^\pm\nu,\tau^\pm\nu$)
\cite{Li:2006nv}.
In Table 1 we show the SM rates and the current
measurements of the purely leptonic decays 
$D^\pm\to l^\pm\nu_l$ and $D^\pm_s\to l^\pm\nu_l$
from CLEO-c.
The displayed measured value of BR($D^\pm_s\to \tau^\pm\nu$)
is an average of separate searches for $\tau^\pm\to \pi^\pm\nu$ and 
$\tau^\pm\to e^\pm\nu\overline \nu$.
For our values for the SM BRs we take 
$f_D=200$ MeV \cite{Aubin:2005ar}
and $f_{D_s}=250$ MeV \cite{Aubin:2005ar},\cite{Juttner:2003ns}. 
The current uncertainty
of $\sim 15\%$ in the lattice calculations of the decay constants
induces an error of $\sim 30\%$ for all leptonic BRs.  
Previous searches for $D^\pm\to \mu^\pm\nu_\mu$
can be found in \cite{Ablikim:2004ry} and a summary of measurements 
for BR($D^\pm_s\to \mu^\pm\nu_\mu,\tau^\pm\nu_\tau$) prior to the
CLEO-c programme is given in \cite{Soldner-Rembold:2001zk}.

In addition to the above charm facilities, the large 
amount of charm data at the B factories can  
be used to search for leptonic decays of $D^\pm$ and $D_s^\pm$.  
\begin{table}
\begin{center}
\begin{tabular} {|c|c|c|c|c|} \hline
Decay & SM BR& Current Exp BR & CLEO-c error &
BES-III error \\ \hline
 $D^\pm\to e^\pm\nu$ & $ 8.24\times 10^{-9}$
& $< 2.4 \times 10^{-5}$ \cite{Artuso:2005ym} & 
 & \\ \hline
 $D^\pm\to \mu^\pm\nu$ & $3.50 \times 10^{-4}$ 
& $4.40\pm 0.66^{+0.09}_{-0.12}\times 10^{-4}$
 \cite{Artuso:2005ym} & $\sim 10\%$
&  $\sim 2\%$\\ \hline
 $D^\pm\to \tau^\pm\nu$ & $9.25\times 10^{-4}$
& $<2.1\times 10^{-3}$ \cite{Rubin:2006nt} & 
& \\ \hline
$D^\pm_s\to e^\pm\nu$ & $1.23 \times 10^{-7}$
& $<3.1 \times 10^{-4}$ \cite{Artuso:2006kz} & 
 &   \\ \hline
 $D^\pm_s\to \mu^\pm\nu$ & $5.22\times 10^{-3}$ 
& $6.57\pm 0.9\pm 0.28\times 10^{-3}$ \cite{Artuso:2006kz} 
& $\sim 10\%$ 
& $\sim 2\%$\\ \hline
 $D^\pm_s\to \tau^\pm\nu$ & $5.09\times 10^{-2}$
& $6.5\pm 0.8$   \cite{Stone:2006cc}
& $\sim 10\%$
& $\sim 1.5\%$\\ \hline
\end{tabular}
\end{center}
\caption{The leptonic decays of $D^\pm$ and $D^\pm_{s}$:
BR predictions in the SM (uncertainty $\sim \pm 30\%$ not shown), 
current measurement from CLEO-c (third column),
expected final precision at CLEO-c (fourth column) and BES-III
(fifth column).}
\end{table}
At BABAR, a search for the process 
$D_s^{\pm *}\to D_s^\pm \gamma\to \mu^\pm\nu_\mu\gamma$ was performed
with 230 fb$^{-1}$ of integrated luminosity, with measured BR 
\cite{Aubert:2006sd}: 
\begin{equation}
{\rm BR}(D_s^\pm\to \mu^\pm\nu)=(6.74\pm 0.83\pm 0.26\pm 0.66)\times 10^{-3}
\end{equation}
The third error (which is not present in the CLEO-c measurement)
is from BR$(D_s^\pm\to \phi\pi^\pm)$, but should decrease with improved 
measurements of BR$(D_s^\pm\to \phi\pi^\pm)$ 
at CLEO-c and forthcoming BES-III.
No search for $D_s^\pm\to \mu^\pm\nu_\mu$ 
has yet been performed by BELLE. In summary, the precision for the 
measurements of BR($D^\pm_s\to \mu^\pm\nu_\mu,\tau^\pm\nu_\tau$) is constantly
improving and  will approach the percent level in the next few years. 
In contrast, 
a significant improvement in the measurement of BR$(B^\pm\to \tau^\pm\nu_\tau)$
would require a luminosity upgrade of the existing B factories.

\section{Numerical Results}

The current measurement of $B^\pm\to \tau^\pm\nu_\tau$ 
(Eq.~(\ref{exp_btaunu}))
restricts $R$ to an interval which can be used to predict 
a region for $r_s$ i.e. the deviation of 
BR$(D^\pm\to \mu^\pm\nu_\mu,\tau^\pm\nu_\tau)$ from the SM rate.
The BELLE collaboration
\cite{Ikado:2006un} uses $V_{ub}=4.39\pm 0.33\times 10^{-3}$
and the unquenched lattice result $f_B=216\pm 22$ MeV 
\cite{Gray:2005ad} which give the theoretical prediction 
BR($B^\pm\to \tau^\pm\nu_\tau)=1.59\pm 0.4\times 10^{-4}$.
Comparing this value for BR($B^\pm\to \tau^\pm\nu_\tau$) 
with the measurement
in Eq.~(\ref{exp_btaunu}) gives
the scale factor $r_H=1.13\pm 0.51$. The interpretation in the
2HDM leads to two allowable regions for $R$. 
The region of larger $R$ is $0.22$ GeV$^{-1}$ $< R < 0.33$ GeV$^{-1}$
for $\tilde\epsilon_0=0$ in Eq.~(\ref{mod_r_H}).  
Note that using smaller values of $f_B$ and $V_{ub}$ for the
SM prediction for BR($B^\pm\to \tau^\pm\nu_\tau)$ would
shift the above allowed region for $r_H$ to larger values.
Therefore for our evaluation of $r_s$
we consider $0.20$ GeV$^{-1}$ $< R < 0.40$ GeV$^{-1}$.
\begin{figure}
\centerline{\includegraphics[width=7cm,height=7cm]{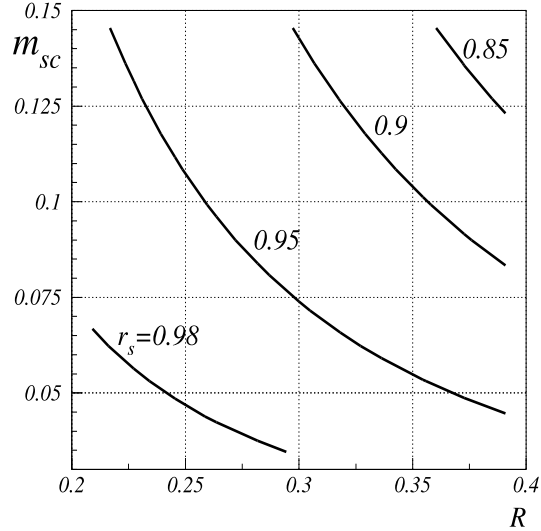}}
\caption{Contours of $r_s$ in the plane $[R,m_{sc}]$, 
where $R=\tan\beta/m_{H^\pm}$ and $m_{sc}\equiv m_s/(m_s+m_c)$.}
\label{fig2}
\end{figure} 
For the quark masses $m_s$ and $m_c$ we use the Particle Data Group
\cite{Yao:2006px} values and obtain 
$0.03 < m_s/(m_c+m_s) < 0.12$ at the $2\sigma$ level.
However, we will consider the range $0.03 < m_s/(m_c+m_s) < 0.15$
which allows for enhancement of the $s$ quark Yukawa coupling
(which is the source of the numerator in $m_s/(m_c+m_s)$)
in the context of the 2HDM (Model III).
In Fig.~{\ref{fig2} we plot contours of $r_s$ in the plane 
$[R, m_s/(m_c+m_s)]$.
The suppression of the SM rate is between $2\%$ and $15\%$ in
most of the parameter space. Thus the $H^\pm$ contribution can be
larger than the expected BES-III precision
($\sim 2\%$) for BR($D^\pm_s\to \mu^\pm\nu_\mu,\tau^\pm\nu_{\tau})$ 
and comparable to the anticipated error in future 
lattice calculations of $f_{D_s}$.
The presence of $H^\pm$ would lead to a deceptively smaller 
{\sl measured} value of the decay constant $f_{D_s}$.
We suggest that the possible 
effects of any $H^\pm$ should not be overlooked when comparing the 
experimentally extracted $f_{D_s}$ to the lattice calculations.
Importantly, discovery of $H^\pm$ in direct production
channels at the Tevatron or LHC with properties corresponding to the
region of $0.20$ GeV$^{-1}$ $< R < 0.40$ GeV$^{-1}$
would necessitate inclusion of the scaling factor $r_s$ 
in BR($D^\pm_s\to \mu^\pm
\nu,\tau^\pm\nu_{\tau}$) in order to do a proper comparison 
of lattice calculations with experiment.

The sizeable error in the decay constant makes it difficult to
confirm the presence of $H^\pm$ from measurements of BR($D^\pm_s\to \mu^\pm
\nu,\tau^\pm\nu_{\tau}$) alone.
An additional observable which is more sensitive 
to the contribution from $H^\pm$ is the
ratio of the muonic decay rates ${\cal R}_{\mu}$ defined by
\begin{equation}
{\cal R}_{\mu}={BR(D^\pm_s\to \mu^\pm\nu_\mu)/BR(D^\pm\to \mu^\pm\nu_\mu)}
\sim r_s(f_{D_s}/f_{D})^2
\end{equation}
The effect of $H^\pm$ on BR$(D^\pm\to \mu^\pm\nu_\mu)$ is negligible
and ${\cal R}_{\mu}$ is proportional to $r_s$. 
The lattice prediction for $f_{D_s}/f_{D}$
is known with greater precision than the prediction for the 
individual values of the decay constants and
hence ${\cal R}_{\mu}$ is a better
probe of the $H^\pm$ effect on BR$(D^\pm_s\to \mu^\pm\nu_\mu)$.
A similar ratio (${\cal R}_{\tau}$) for the tauonic decays could be
measured at BES-III \cite{Li:2006nv}
(and possibly at CLEO-c) although we only consider 
${\cal R}_\mu$ for which both experiments expect superior precision. 
An unquenched lattice calculation \cite{Aubin:2005ar} gives 
$f_{D_s}/f_D=1.24\pm 0.07\pm 0.01$.
The first measurements of $f_{D_s}/f_D$ from
CLEO-c \cite{Stone:2006cc} and BABAR
\cite{Aubert:2006sd} are $1.26\pm 0.11\pm 0.03$ and $1.27\pm 0.14$, where
both use the CLEO-c result for $f_{D}$ \cite{Artuso:2005ym}.
The central values are in agreement with the above 
unquenched lattice result, and
differ from the quenched results which typically give $f_{D_s}/f_D=1.10$
\cite{Ryan:2001ej}.

In Fig.~\ref{fig3} the ratio ${\cal R}_\mu$ is plotted as a function of
$R$ for $m_s/(m_s+m_c)=0.08$. We 
take $f_{D_s}/f_D=1.24\pm 0.02$
which might correspond to the error in the lattice calculations
as BES-III approaches $2\%-3\%$ precision for these decays. 
For $f_{D_s}/f_D=1.24$ one can see
that ${\cal R}_\mu$ varies from 14.7 to 13.3 
(i.e. a $10\%$ suppression) as $R$ varies from $0 \to 0.4$ 
GeV$^{-1}$. Hence precise measurements of ${\cal R}_{\mu}$ 
could favour or disfavour $H^\pm$.

\begin{figure}
\centerline{\includegraphics[width=7cm,height=7cm]{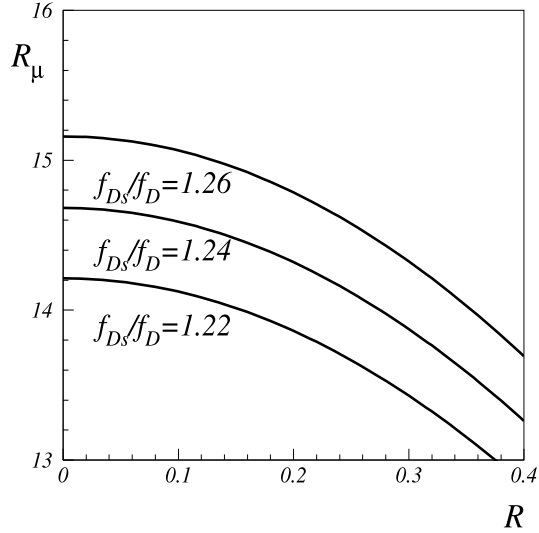}}
\caption{${\cal R}_{\mu}$ as a function of 
$R(=\tan\beta/m_{H^\pm})$ for $m_s/(m_s+m_c)=0.08$
and $f_{D_s}/f_D$=1.22, 1.24 and 1.26.}
\label{fig3}
\end{figure}

\section{Conclusions}
The recent observation of $B^\pm\to \tau^\pm\nu$ does not 
preclude a sizeable contribution from $H^\pm$ in the 2HDM or MSSM. 
Such a $H^\pm$ would contribute to the leptonic decays
$D^\pm_{s}\to \mu^\pm\nu_\mu,\tau^\pm\nu_{\tau}$. 
We showed that $H^\pm$ can suppress
BR($D^\pm_{s}\to \mu^\pm\nu_\mu,\tau^\pm\nu_{\tau}$) by 
$10\%$ or more, which is larger than the expected experimental error 
($2\%\to 3\%$) for future measurements of these decays.
We emphasize that new physics effects like these 
should not be overlooked when
comparing the experimental measurements of $f_{D_s}$
to the lattice predictions. Moreover,
favouring or disfavouring the presence of $H^\pm$ could be obtained
by precise measurements of the ratio 
${\rm BR}(D^\pm_s\to \mu^\pm\nu_\mu)/{\rm BR}(D^\pm\to \mu^\pm\nu_\mu$).
Prospects for $D^\pm_s\to 
\mu^\pm\nu_{\mu}$ and $D^\pm_s\to\tau^\pm\nu_{\tau}$
are bright with ongoing CLEO-c and forthcoming BES-III,
and we encourage their study together with $B^\pm\to \tau^\pm\nu$
at the B factories.

\section*{Acknowledgments}
Comments from Sheldon Stone are gratefully acknowledged. 
A.G.A was supported by National Cheng Kung University grant
OUA 95-3-2-057.

\end{document}